\documentclass[11pt,twoside]{article}
\usepackage{asp2010}

\resetcounters

\usepackage{natbib}
\begin{document}

\title{Metallicity Gradients in Simulated Disk Galaxies}
\author{Kate Pilkington$^{1,2,3}$ and Brad K. Gibson$^{1,2,3}$
\affil{$^1$Jeremiah Horrocks Institute, UCLan, Preston, PR1~2HE, UK}
\affil{$^2$Monash Centre for Astrophysics, Monash Univ, Clayton, 3800, Australia}
\affil{$^3$Dept of Physics \& Astronomy, Saint Mary's Univ, Halifax, B3H~3C3, Canada}
}

\begin{abstract}
The stellar metallicity and abundance ratio gradients from the fiducial 
late-type galaxy simulation of \citet{Stinson10} are presented. Over 
$\sim$1$-$3 scalelengths, gradients are shown to 
flatten with time, consistent with empirical evidence at high- and
low-redshifts.  Kinematic effects, including radial migration,
though, flatten these intrinsicly steep gradients such that by redshift 
$z$=0, the measured gradients of these (now) old stars are flatter than 
their young counterparts, in contradiction to what is observed locally. 
Conversely, the stellar [O/Fe] gradient is (to first order) robust 
against migration, remaining fairly flat for both young and old 
populations today.
\end{abstract}

\vspace{-5.0mm}
\section{Radial Gradients} 
\vspace{-2.0mm}
We examine the abundance gradients for 25 
cosmological simulations and 2 Milky Way
chemical evolution models, to quantify the 
impact of hydrodynamic algorithms (e.g. SPH vs AMR), star 
formation thresholds, and energy feedback, in 
establishing gradients within the inside-out paradigm of galaxy 
formation \citep{Pilkington12}.

We extend our work, using the fiducial disk (g15784) from 
\citet{Stinson10}, to determine its abundance \it ratio \rm 
gradient.  Disk stars are identified using the \citet{Abadi03}
kinematic decomposition; the disk
scalelength is $\sim$3~kpc. The star formation history of
the kinematically-defined component is 
shown in Fig~1.  The late-time behaviour can be 
characterised by an exponential declining with an $\sim$7~Gyr timescale 
to a present-day rate of $\sim$1$-$2~M$_\odot$/yr), consistent with the 
Milky Way.

\begin{figure}[!ht] 
\centering
\includegraphics[scale=.6]{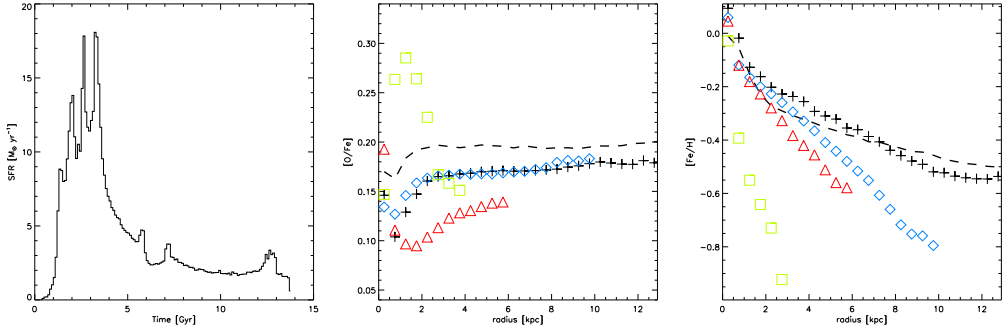} 
\caption{\it Left\rm: Star formation history of the 
kinematically-defined disk of g15784; \it Middle\rm: Radial [O/Fe] 
gradients for the young (born in the most recent 0.1~Gyr) disk stars at 
three redshifts: $z$=0.0 (plus symbols); $z$=0.5 (diamonds); $z$=1.0 
(triangles); $z$=2 (squares). 
Right\rm: Radial [Fe/H] gradients for the same `young' 
stars at the same three redshifts (symbols as for the middle panel).
The dashed curve in the latter panels is the mass-weighted gradient
using all stars born since $z$=2.}
\label{Grad} 
\end{figure} 

In the middle \& right panels of Fig~1, radial [O/Fe] and [Fe/H] 
gradients are shown for young stars at four redshifts,
where, by `young', we mean those 
corresponding to OB-stars at these redshifts. At early epochs, 
recently-formed stars show steeper [O/Fe] and (especially) [Fe/H] 
gradients; for the latter, young stars at high-$z$ show an [Fe/H] 
gradient of $-$0.08~dex/kpc, compared with the young stars at $z$=0 
(which show $-$0.04~dex/kpc).  As shown by \citep{Pilkington12}, this is 
consistent with the \it direct \rm measure of steeper gas-phase 
gradients at high-$z$ by \citet{Yuan11}, as well as the \it indirect \rm 
measure of steeper gradients at early times \it inferred \rm from the 
steeper gradients of older planetary nebulae \citep{Maciel03} and open 
cluster \& field star giants \citep{Yong06}, relative to younger 
tracers.

This latter conclusion merits further discussion; specifically, when the 
same star particles are viewed today (mimicing in a \it direct \rm sense 
the manner in which \citet{Maciel03} and \citet{Yong06} inferred the 
above conclusions regarding early-time gradients from present-day data) 
- i.e., via the classification of 'old' vs 'young' star at $z$=0, one 
finds the apparent opposite conclusion (that old tracers today possess 
shallower abundance gradients than young tracers, similar to what was 
found by \citet{San09}).  This apparent discrepancy is a reflection of 
the fact that the older (steeper [Fe/H] gradient) star particles have 
preferentially migrated/scattered to larger galactocentric distances, 
relative to the younger star particles (shallower [Fe/H] gradients).  
Binned to match \citet{Maciel03}, we find that 6$-$9~Gyr old stars show 
a flattening in their gradient by $z$=0 of $\sim$0.05~dex/kpc, inverting 
the situation shown in Fig~1 such that old stars in the simulations have 
shallower gradients than the younger stars.  Importantly, this does not 
impact on gas-phase abundance gradient analyses.

Interpreting the stellar [O/Fe] gradients is less complicated by these 
migrations; since $z$$\sim$1.0, stars are born with mildly
inverted (but essentially flat) 
[O/Fe] gradients (with gradients $<$$+$0.005~dex/kpc). At $z$$\sim$2, 
the compact disk is apparent (middle panel of Fig~1); the $\alpha$-enhanced
disk stars in the inner $\sim$2.5~kpc, \it by redshift $z$$\sim$0\rm,
are radially distributed (via migration/kinematic heating) throughout
the disk, yielding (again) very flat [O/Fe] gradients over 
$\sim$1$-$3 disk scalelengths. Over the same radial range, these
flat gradients are consistent with those observed in nearby disk
galaxies \citep{San11} and in the Milky Way \citep{Yong06}.

\vspace{-1mm}
\acknowledgments 
We acknowledge the support of the UK's STFC and the financial 
assictance of Monash and Saint Mary's University, and the Organizing
Committee of the 3rd Subaru International Conference on Galactic 
Archaeology.

\bibliography{Pilkington}

\end{document}